\documentclass{emulateapj}

\shorttitle{Infrared Emission from Abell 2597}
\shortauthors{Donahue et al.}
\usepackage{graphics,graphicx,epsf,natbib}
\received{}

\begin{document}
\title{Infrared Emission from the Nearby Cool Core Cluster Abell 2597}
\author{
Megan Donahue\altaffilmark{1},
Andr\'es Jord\'an\altaffilmark{2},
Stefi A. Baum\altaffilmark{3},  
Patrick C\^ot\'e\altaffilmark{4},
Laura Ferrarese\altaffilmark{4}, 
Paul Goudfrooij\altaffilmark{5}, 
Duccio Macchetto\altaffilmark{5}, Sangeeta Malhotra\altaffilmark{6}, 
Christopher P. O'Dea\altaffilmark{7}, 
James E. Pringle\altaffilmark{8,5}, James E. Rhoads\altaffilmark{6}, 
William B. Sparks\altaffilmark{5}, G. Mark Voit\altaffilmark{1} }

\altaffiltext{1}{Michigan State University, Physics \& Astronomy Dept., East Lansing, MI 48824-2320, donahue@pa.msu.edu}
\altaffiltext{2}{European Southern Observatory, Karl-Schwarzchild-Stra{\ss}e 2, 85748 Garching ber M\"unchen, Germany, ajordan@eso.org}
\altaffiltext{3}{Carlson Center for Imaging Science, Rochester Institute of Technology, Rochester, NY}
\altaffiltext{4}{Herzberg Institute of Astrophysics, National Research Council of Canada,  
5071 West Saanich Rd, Victoria, BC V8X 4M6, Canada}
\altaffiltext{5}{Space Telescope Science Institute, 3700 San Martin Drive, Baltimore, MD 21204}
\altaffiltext{6}{Physics Department, Arizona State University, Tempe, AZ 85287-1504}
\altaffiltext{7}{Physics Department, Rochester Institute of Technology, 84 Lomb Memorial Dr, Rochester NY, 14623-5603}
\altaffiltext{8}{Institute of Astronomy, Madingley Road, Cambridge, UK}

\begin{abstract}
We observed the brightest central galaxy (BCG) in the nearby ($z=0.0821$) cool core 
galaxy cluster Abell 2597 with the  
IRAC and MIPS instruments on board the Spitzer Space Telescope.
The BCG was clearly detected in
all Spitzer bandpasses, including the 70 and 160 $\mu$m wavebands. We
report aperture photometry of the BCG. The spectral energy
distribution exhibits a clear excess in the FIR over a Rayleigh-Jeans
stellar tail, indicating a star formation rate of $\sim 4-5$ solar masses 
per year, consistent with the estimates from the UV 
and its H$\alpha$ luminosity. 
This large FIR luminosity is consistent with that of a starburst or a Luminous Infrared Galaxy (LIRG), but together
with a very massive and old population of stars that dominate the energy output of
the galaxy. If the dust is at one temperature, the ratio of 70 to 160 micron fluxes indicate that the 
dust emitting mid-IR in this source is somewhat hotter than the dust emitting mid-IR in two 
BCGs at  higher-redshift ($z\sim0.2-0.3$) and higher FIR luminosities 
observed earlier by Spitzer, in clusters Abell 1835 and Zwicky 3146.

\end{abstract}


\section{Introduction}

The brightest cluster galaxies (BCGs) in the cores of the most X-ray luminous clusters of galaxies
are the most massive galaxies in the universe.  Their unusually extended stellar 
envelopes, high optical luminosities, and red colors 
pose a challenge to models of galaxy formation, which, conversely, predict
that the brightest cluster galaxies should be even more luminous than observed 
and should be blue, not red \citep{2004MNRAS.347.1093B}.  According to such models,
galaxies at the centers of clusters should be highly luminous and blue because supernova 
feedback is unable to prevent runaway cooling, condensation, and star formation in the gas 
at the center of a cluster.  However, even though the galaxies at the centers of X-ray luminous 
clusters are red, Spitzer Multiband Imaging Photometer for Spitzer (MIPS) 
imaging of moderate redshift ($z\sim 0.2-0.3$)
clusters with cool cores and large H$\alpha$ luminosities 
has revealed that they can still be surprisingly luminous 
infrared sources, $\sim 10^{44-45}$ erg s$^{-1}$ \citep[][]{2006ApJ...647..922E}. 
Their mid-infrared luminosity turns out to be $\sim0.1-1\%$ times the total 
X-ray luminosity from the
hot intracluster medium (ICM), and puts these galaxies in the same luminosity class 
as LIRGs (Luminous Infrared Galaxies.)

Historically, clusters in which the central gas radiates enough energy to cool and condense within
a Hubble time have been categorized as cooling-flow clusters.  Such clusters exhibit highly peaked 
X-ray fluxes and, usually, ICM temperature gradients that decrease
into the cluster core. Cool core clusters are rather common at $z<0.4$, accounting for $\sim50\%$ of
the X-ray luminous population,  indicating that this phase is not short lived. Yet, the central
gas in these clusters does not appear to be cooling and condensing at the rates implied by 
naive interpretations of X-ray imaging data, which could exceed 100 M$_{\odot} \, {\rm yr}^{-1}$.  
Central star formation rates are much lower than this, and the central galaxies do not appear to 
contain the $\sim 10^{12} M_\odot$ of cold gas that should have collected over a 
Hubble time  \citep[eg.][]{1994ARA&A..32..277F}.  This was the
notorious  ``cooling flow'' problem:  How can these systems be so common, when 
there is no obvious source of heat to replenish the energy these systems radiate so quickly?

High-resolution X-ray spectroscopy has given the cooling-flow problem an additional twist
by showing that most of the X-ray gas with a short cooling time does not even cool 
below  $10^7$K  \citep[e.g., ][]{2001A&A...365L.104P}.  
Hot gas is detected at a range of temperatures down to $\sim 1/3$ of the virial temperature but does 
not appear to emit at lower temperatures. This result confirms that the gas condensation rates in these systems are not very large but doesn't explain what suppresses cooling.
Recent Chandra X-ray observations have suggested that AGN feedback might be what
 inhibits cooling in cluster cores, because some (but not all)
cooling flow clusters host central AGN which excavate kpc-scale cavities in the ICM 
 \citep[eg.][]{2001ApJ...562L.149M}. Still more clusters exhibit elevated
central entropy levels which may have been produced by a now-inactive 
AGN  \citep{Donahue2005A, 2006ApJ...643..730D}. 
There is increasing support for the idea that the central
AGN is the culprit providing the ``extra" heating, not only countering radiative cooling
in cluster cores, but also providing a self-regulating feedback mechanism in massive galaxies
in general \citep[e.g.,][]{2006MNRAS.368L..67B}.
These observations and theories have sparked a renewed interest in the role of AGN
in the formation of galaxies, and in particular, their role in heating the ICM and stifling star
formation.

Interestingly, the mechanism that counteracts uninhibited radiative 
cooling in cluster cores does not appear to shut
off star formation altogether.  Cooler gas is not completely absent in the central galaxies, 
a few of which have substantial 
quantities ($\sim 10^{10-11}$ M$_\odot$) of molecular gas at a range of temperatures.
Extended, vibrationally-excited H$_2$ 2-micron emission lines 
have been detected in these sources
 \citep[][]{1994iaan.conf..169E, 2001MNRAS.324..443J,2006ApJ...652L..21E}, and
about $10^{10-11}$ M$_\odot$ of cold molecular hydrogen has been inferred
from CO observations in many of these systems 
 \citep[e.g., ][]{2001MNRAS.328..762E, 
2003ApJ...594L..13E, 2006A&A...454..437S}.
If some of the ICM cools and forms stars, that could explain why most cool core systems exhibit
luminous optical emission-line nebulae, which in turn nearly always accompany the
presence of molecular gas. Such nebulae produce bright H$\alpha$ and 
forbidden line emission  \citep[][]{1989ApJ...338...48H}.  
Emission-line spectroscopy suggests that
at least some of the emission can be explained as the result of photoionization and heating
by hot stars \citep[][]{1997ApJ...486..242V}. 
Tantalizing possible detections by FUSE of million degree gas, via OVI emission, suggest that
at least some of the gas does cool below X-ray emitting temperatures \citep{2001ApJ...560..187O,
2006ApJ...642..746B}. However, the interpretation of such spectra is complicated, sometimes by the
presence of a low-luminosity AGN and possibly by slow shocks or heating by
cosmic rays or local X-rays.  

In this paper, we present Spitzer observations of mid-infrared emission from the BCG in Abell 2597, a nearby cool-core cluster, that address an important question about the cool-core
phenomenon: what is the dusty star formation rate in this cluster?  These data also
stimulate new questions about the star formation in BCGs:
(1) Is the infrared emission from a BCG in a cool-core cluster typical of a starburst in a spiral galaxy? (2) Is the dust spectrum characteristic of relatively unprocessed 
Milky Way-type dust that has been 
transported to the center of the BCG during a merger, or is it consistent with the dust having been recently exposed
to the harsh environment of the intracluster medium? (3) If a modest amount of gas is condensing from the ICM, is the condensation rate consistent with the observed star-formation activity and molecular gas content?    Our Spitzer observations show that the IR spectrum of this BCG is fairly typical of a normal starburst, with a far-IR peak at $\sim$70$\mu$m and a mid-IR excess consistent with emission from polycyclic aromatic hydrocarbons (PAHs).  The star formation rate implied by the IR and UV emission from this galaxy are also consistent with observational limits on the cooling and condensation rate of gas from the central ICM.  In our analysis, we assume $H_0=70$ km s$^{-1}$ Mpc$^{-1}$,
$\Omega_M=0.3$, and $\Omega_\Lambda=0.7$ cosmology.  At the redshift of the BCG in
A2597 ($z=0.0821\pm0.0002$, \cite{VD1997})  the scale is 1.551 kpc arcsec$^{-1}$.

\section{The Brightest Cluster Galaxy in Abell 2597}

The brightest cluster galaxy in Abell 2597 (Abell richness class 0) 
contains a well-studied FRI radio source PKS2322-12 and 
is conveniently positioned in region of the sky with very low Galactic extinction 
($A_B = 0.131$ mag; 
\citet{1998ApJ...500..525S}), confirmed by
\citet{2003MNRAS.343..315B}.
 The optical spectrum of the central galaxy was studied intensively by \citet{VD1997}, who placed the first
model-independent reddening, temperature, and metallicity constraints on a cluster emission-line nebula using faint forbidden lines. They also determined that the excitation mechanism of the 
emission lines could not be shocks. The preferred excitation source was stars, plus an additional, unidentified, source of
heat. This conclusion was supported by analysis of the infrared emission line spectrum of
Paschen alpha lines and vibrationally excited H$_2$ \citep{2005MNRAS.360..748J}. One slight difference
between these analyses is that the \citet{VD1997} study also placed limits on the UV spectral shape, from the
lack of a measurable He II recombination line. 
 
\citet{Cardiel1998} found radial gradients in the 4000 \AA~ break and Mg$_2$ indices of Abell 2597's BCG, indicating
recent (0.1 Gyr) star formation. \citet{1999ApJ...518..167M} found that its U-band light is unlikely to be
from scattered AGN light, based on polarization limits on the continuum. Most
recently, it has been studied in the far UV using Hubble Space Telescope
STIS observations \citep{ODea2004} and WFPC2 blue and emission-line images \citep{Koekemoer1999}.
Vibrationally excited molecular hydrogen was discovered to trace the same features as the 
optical emission lines by \citet{Donahue2000}.The Chandra X-ray
Observatory has reveals two-sided radio bubbles, surrounded by hot ICM \citep{McNamaraA2597_2001}. 
This BCG also has one of the most convincing FUSE detections of OVI \citep{2001ApJ...560..187O}. Further, $\sim4\times10^9$
M$_\odot$ of cold H$_2$ has been inferred from CO detections \citep{Edge2001}.

The most recent X-ray observations of A2597 are from a 120-kilosecond XMM observation. Analysis of both EPIC and RGS spectroscopy by \citet{2005MNRAS.358..585M} 
suggests a classical cooling flow at the level of $90\pm15$ solar masses per year in the
central region, dropping to about 20 M$_\odot$ yr$^{-1}$ in the central 40 kpc 
surrounding the BCG. This result is based on the joint analysis of vanishingly weak ($\sim 1-2\sigma$) FeXVII features in the RGS and the extremely difficult interpretation of soft X-ray excess in the CCD spectrum. 
While this result at best is only suggestive of cooling, it makes the BCG in Abell 2597 
the only source with published 
emission-line detections from gas at $10^{5-7}$K, suggesting a possible connection between  condensation from the 
hot gas and star formation. 

\section{Observations and Data Reduction}

\subsection{Infrared Array Camera}
The Infrared Array Camera (IRAC) has four wavelength channels, 3.6, 4.5, 5.8, and
8 $\mu$m \citep{2004ApJS..154...10F}.   The Astronomical Observing Request (AOR) number 
for the IRAC observation was 13372160. The total observing time with all 4 detectors
available for IRAC was 270 minutes. Each frame was 100 seconds. Thirty six positions were
dithered for a total of 3600 seconds per bandpass.

We have used the flux-calibrated images from the SSC IRAC pipeline (software version S14.0.0)  for our analysis. The standard pipeline subtracts dark current based on laboratory measurements and a dark-sky frame based on observations of the darkest, star-free parts of the sky.  A flat field, based on observations of zodiacal background and cleaned of cosmic rays, is then divided out of the data.  Finally, the data are flux calibrated, producing images with flux units of MJy per steradian. 
See the IRAC data handbook\footnote{\url{http://ssc.spitzer.caltech.edu/data/hb}} and the IRAC calibration
paper of \citet{2005PASP..117..978R} for more detail.

\subsection{Multiband Imaging Photometer}
The Multiband Imaging Photometer (MIPS) has three bands, with weighted wavelength 
averages of 23.68 $\mu$m, 71.42 $\mu$m, and 155.9 $\mu$m. For convenience, we will
refer to these bands as 24, 70, and 160 $\mu$m. The FWHM of the PSF in those bands 
is $6\arcsec$, $18\arcsec$, and $40\arcsec$ respectively. Our total observing time with 
the MIPS was 36 minutes (total exposure time of 10 seconds per pixel at 24 $\mu$m and 
15 seconds per pixel for 70 and 160 $\mu$m each). The AOR  number 
for the MIPS observation was 13371904. 

\begin{figure}[ht]

\includegraphics[width=180mm]{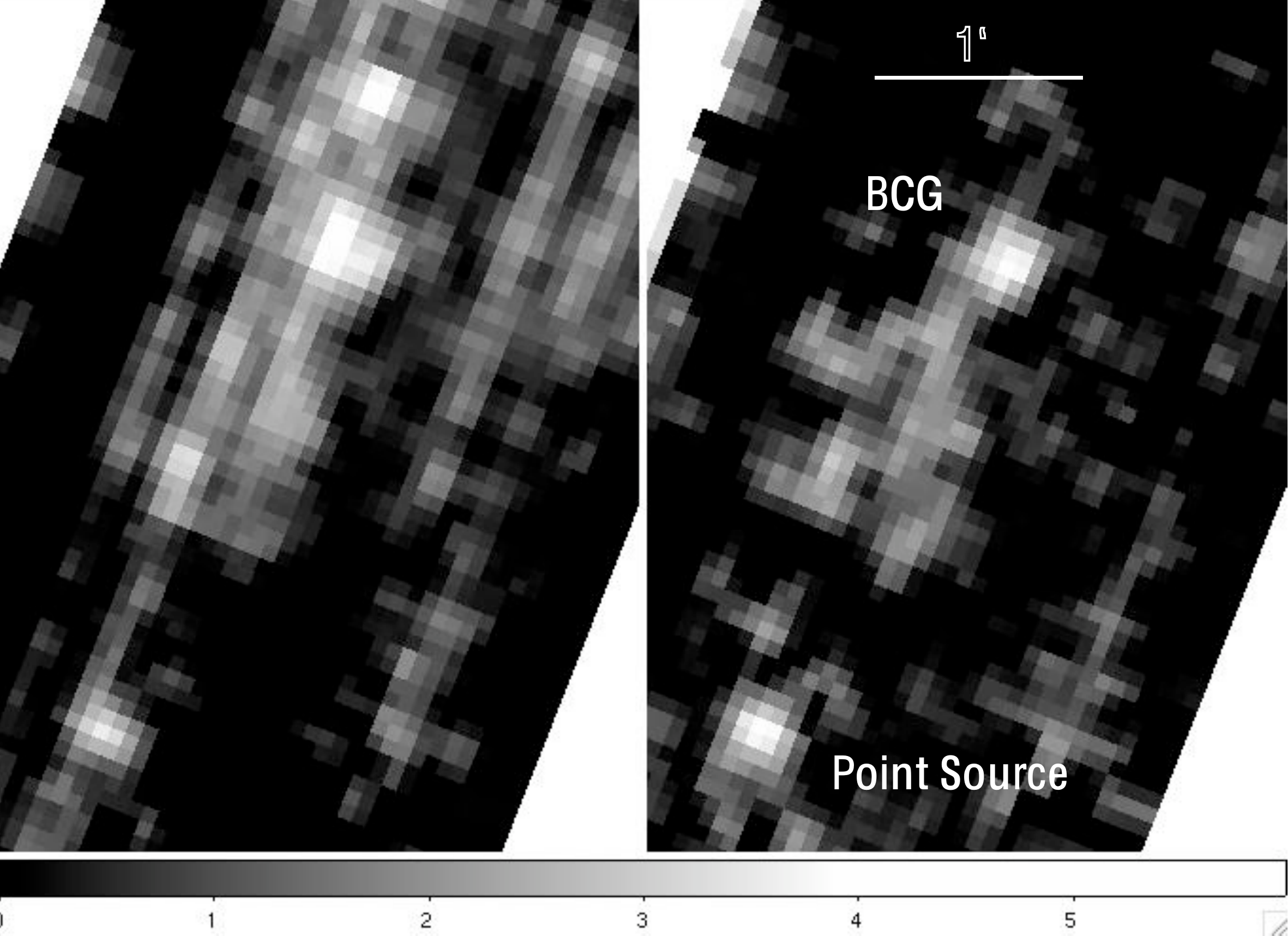}
\caption{
This grey-scale figure shows two matched versions of the MIPS 70 micron image of the brightest
cluster galaxy (BCG) in Abell 2597. North is towards the top of the image, and East is
towards the left. A one-arcminute scale bar is displayed. The units on the grey scale color bar at
the bottom of the figure are MJy per steradian. The left image is of the standard pipeline product,
and the right image shows the same object and data, where we subtracted median sky images from the 
individual exposures using GeRT routines (see text for details), then we co-added
using MOPEX. Only the BCG and the point source in the lower left
hand corner were masked during this procedure. The result was a cleaner image, and an
intriguing hint of an extended 70-micron source, extending approximately 1 arcminute 
south-east of the brighter compact source near the center of the BCG. This feature is likely a residual of the stripes in the original image, as discussed in the text. A point source $1\arcmin$ north of the BCG in the
original image vanishes in the filtered image.
\label{figure:70}}
\end{figure}

We used the standard Spitzer Science Center (SSC) pipeline processing (software version S14.4.0) 
of the 24,  70, and 160 $\mu$m data. 
We investigated whether the 70 and 160 $\mu$m data would benefit from additional work. 
We reprocessed the raw 70 and 160 $\mu$m data with the 
GeRT software package\footnote{GeRT is available from the SSC at \url{http://ssc.spitzer.caltech.edu/mips/gert/}}, following the algorithms derived by the MIPS Instrument 
Team and the MIPS Instrument Support team \citep{2005PASP..117..503G}. We time-filtered
the resulting 70 $\mu$m basic data by subtracting a smoothed version of
the signal obtained with a boxcar median filter of 30 frames.
This procedure subtracts the residual 
time-variations of the response. We then used the MOPEX package\footnote{MOPEX:
MOsaicking and Point source EXtraction, available from the SSC at 
\url{http://ssc.spitzer.caltech.edu/postbcd/download-mopex.html}.} 
to co-add the filtered images. In Figure~\ref{figure:70} we 
show the GeRT-filtered 70 $\mu$m image side by side with the standard pipeline image. Here we note that 
there is some evidence for a faint extended region in the filtered data. However, since this region is
aligned with the higher-noise streak in the pipeline image, subtracted from the GeRT data, 
and since a mosaic of the individual pipeline-filtered exposures does not show this
feature, we do not make strong claims about its reality. 

We reduced the 160 $\mu$m data in a very similar way, using GeRT.
However, since there are very few pixels and the central source is marginally extended in an
odd wedge-shape, additional filtering in the time domain did not result in significantly cleaner 160 $\mu$m images. 
The exclusion of individual images taken directly after the 
periodic ``stim" images had no effect on the final product.  
The MOPEX package was used to combine the 160 $\mu$m basic calibration products.

\subsection{Data Analysis}

We measured the flux in all seven wavebands 
within a circular aperture of 
$r=25\arcsec$ ($\sim39 h_{70}^{-1}$ kpc), centered on 
the coordinates $\alpha=351.33199$ and $\delta = -12.124612$ (J2000). 
(RA of $23^h$ $25^m$ $19.6^s$, DEC of -12 07' 29").  
The background was estimated from pixels in an annulus between 
$38.5\arcsec$ and $42.5\arcsec$ 
from the center. We also measured the flux in a larger aperture ($r=35\arcsec$)
for the MIPS 70 and 160 $\mu$m images. 
A histogram of the 
background values was fit to a Gaussian initially centered on the median
background per pixel.
The mean background value was subtracted from the aperture
flux. The net fluxes were averaged  and multiplied by the aperture sky
area in steradians to yield the flux in Janskys. 
No aperture correction was required for the IRAC photometry.

Since the MIPS images are nearly point sources, we used the APEX package to 
fit point response functions (PRFs) and obtain total fluxes for the MIPS. For comparison, we
report both the large aperture flux and the PRF flux for each band. We obtain reasonable agreement
except in the case of the 160 $\mu$m aperture flux, for which the aperture correction,
even at $r=35\arcsec$ we estimate to be $\sim1.5$, based on the convolution of the
standard PRF with the source. The
IRAC flux uncertainties are $\sim5\%$. The
color terms at 3.6 and 4.5 $\mu$m are unlikely to be significant; however, if
the spectrum is dominated by PAHs at 5.8 and 8 $\mu$m, the color-term corrections
could be significant, up to $\sim50\%$. MIPS absolute flux uncertainties are 10\% at 
24 $\mu$m and 20\% at 70 and 160 $\mu$m. We report raw and aperture-corrected 
aperture fluxes in Table~\ref{table:fluxes}.

\begin{figure}[ht]
\includegraphics[width=180mm]{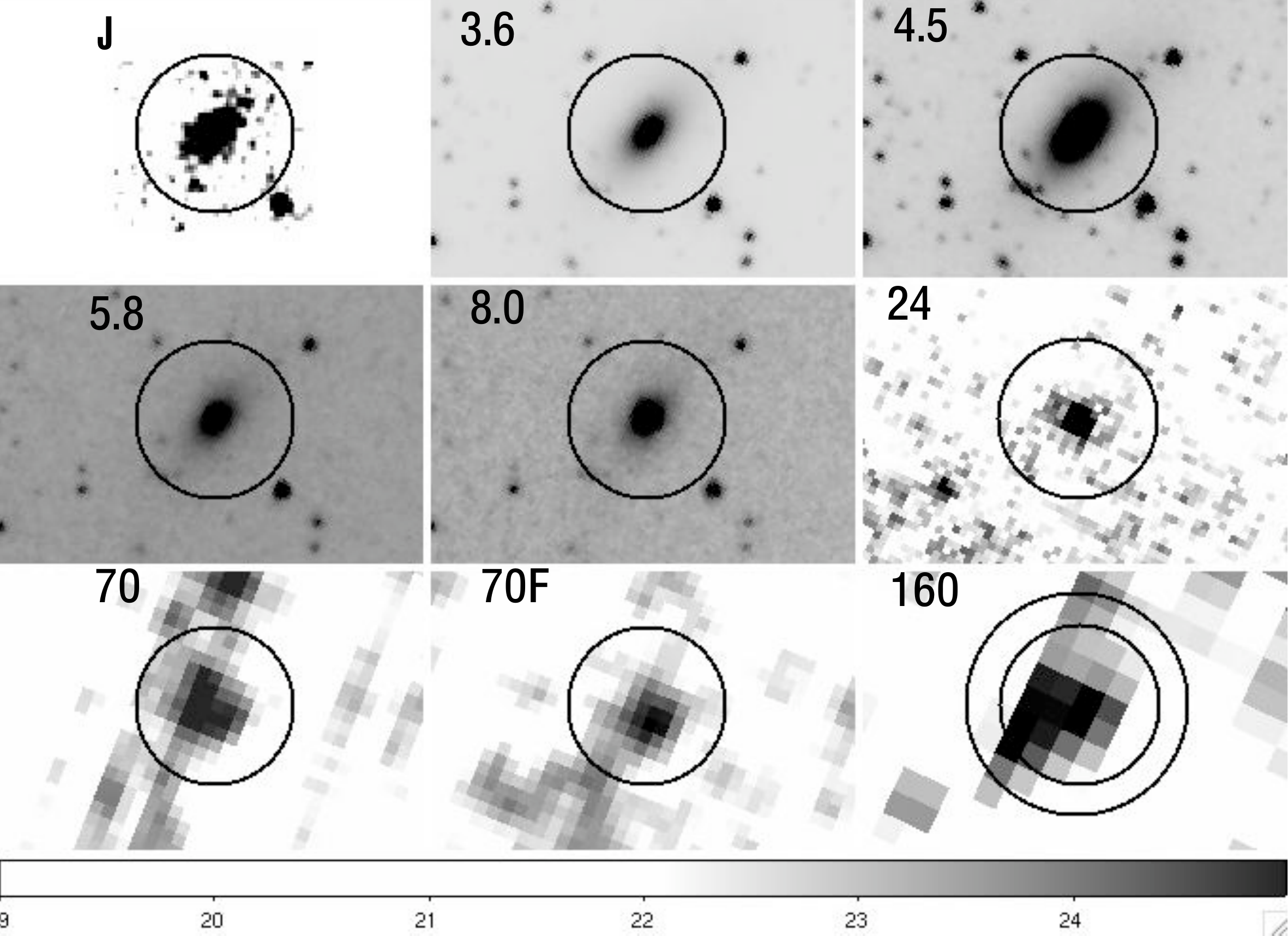}
\caption{This grey-scale figure shows matched multiwavelength infrared images of the brightest cluster
galaxy in Abell 2597. The angular scale of each image is $130\arcsec$ horizontal and $86\arcsec$
vertically. North is up and East is to the left. All subimages have the same angular scale. 
Left to right, top row: J-band from 2MASS, IRAC 3.6 and 4.5 microns.
Second row: IRAC 5.8, 8.0 microns, MIPS 24 microns. Third row: MIPS 70 micron image from the
SSC pipeline, the median-sky-subtracted 70 micron image (70F), and the MIPS 160 micron image.
An $r=25\arcsec$ aperture is plotted for scale over each subimage. The 160 micron image also
shows an $r=35\arcsec$ aperture. \label{figure:all}}
\end{figure}

We extracted photometry for this galaxy from the 
Two Micron All Sky Survey (2MASS) extended source 
catalog \citep{2000AJ....119.2498J}\footnote{Described in \url{http://spider.ipac.caltech.edu/staff/jarrett/2mass/XSC/}
and query service available through GATOR \url{http://irsa.ipac.caltech.edu/applications/Gator/}}
in J, H, and K infrared bands. 
Absolute photometric calibration from \citet{2003AJ....126.1090C} converts 
the 2MASS magnitudes to Janskys.
The 2MASS total apertures (the apertures that measured the total light from the extended source) 
of $25.35\arcsec$ were  similar to that used for the Spitzer data.  We show all images (2MASS and Spitzer)
in Figure~\ref{figure:all} together with the $r=25\arcsec$ aperture in the 1-24 micron images and
the $r=25\arcsec$ and $35\arcsec$ aperture for the 160 micron image, for scale. 

\section{Discussion}

The total amount of far-IR emission from the BGC in Abell 2597 is  
$\nu L_\nu \sim 1 \times 10^{44}$ erg s$^{-1}$, corresponding to $\sim 4 M_\odot \, {\rm yr}^{-1}$ according to \citet{Kennicutt1998}.  This  
FIR luminosity is as high as that  
of a  LIRG (Luminous Infrared Galaxy), and the broad-band spectrum is consistent with 
that of a starburst.  
The IR-inferred star formation rate is also consistent with the conclusion 
of \citet{ODea2004} that hot stars, detected in the UV, forming at the rate of a few solar masses per year, are the dominant source of ionization of the optical nebula.  The emission we detect with IRAC is extended, but it appears to be mainly associated with the central galaxy. The mid-IR sources,
except at 24 $\mu$m, are not well-fit by point sources, but those  too are 
completely contained within the boundaries of the stellar light of the BCG.
Because of Spitzer's 
diffraction limit, we cannot say much about its mid-IR morphology.  It is also very difficult to place any 
flux limits on extremely extended emission, beyond the galaxy, such as emission 
from the cluster ICM itself, due to the nature of mid-IR observations which require
multiple, offset exposures to assess the background and the current 
state of knowledge about absolute Spitzer backgrounds.  Because of the positional 
association of the Spitzer-detected IR 
emission with the BCG, we suspect that what we detect is entirely interstellar, not intracluster, in nature.

\begin{figure}[ht]

\includegraphics[width=180mm]{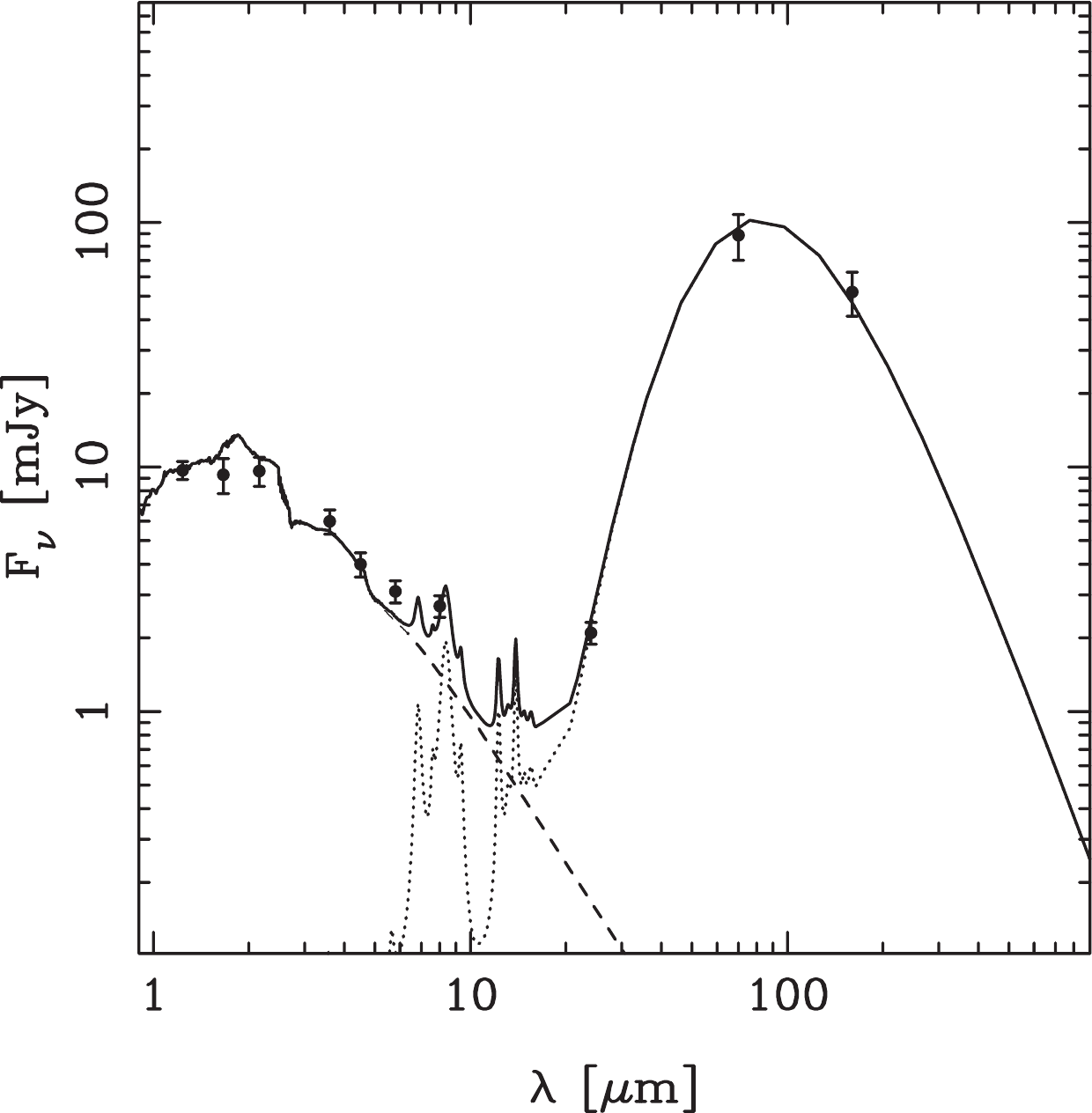}
\caption{ \label{figure:SED}
Spectral energy distribution (SED) of the central galaxy of Abell~2597. 
The solid symbols indicate the 
2MASS and \textit{Spitzer} IRAC and MIPS measurements presented in 
Table~\ref{table:fluxes}.
The error bars include systematic uncertainties as 
described in the text. The solid line is the best-fit 
two component SED model composed of a giant elliptical SED taken from
the SED template library of the Hyper-z photometric redshift code (dashed line) 
\citep{2000A&A...363..476B}
and a nuclear starburst SED model (dotted line)
from \citet{2007A&A...461..445S}. The inferred mass of the giant
elliptical is $\sim 3.12 \times 10^{11} ~\rm{M}_\odot$
The best-fit component from the
Siebenmorgen \& Kr\"ugel library  is indicated with a dotted line,
and corresponds to a mid-IR starburst nucleus of $r=0.35$~kpc where 60\% of the 
luminosity is from hot spots, dense clouds ($n=10^2$ cm$^{-3}$) 
around buried OB stars, with a visual extinction $A_V=36$ mag. 
The total luminosity of this component is $L=10^{10.4}\,\, L_{\odot} \sim 0.95 \times 10^{44}$ 
erg sec$^{-1}$.
}
\end{figure}

Figure~\ref{figure:SED} compares the broad-band spectral energy distribution of this galaxy with the 
spectrum expected from an old stellar population with zero dust.  The near-infrared spectrum 
is fairly flat, as one would expect from an old stellar population at this redshift, and the 
stellar mass implied by the near-IR luminosity is $\sim 3.12 \times 10^{11}$ M$_\odot$. 
Stellar continuum emission is expected to dominate in the 3.5 and 4.5 $\mu$m bands, and 
the ratio of those two bands to the near-IR emission is consistent with that expectation.  Those  bands are well-fit by a simple giant elliptical spectral template, obtained from the SED template 
library of the Hyper-z photometric redshift code \citep{2000A&A...363..476B}.   In the far-IR, one can see the prominent 70 - 160 $\mu$m peak observed with MIPS.  Such a peak is expected from dust 
in active star-forming regions.   Excess emission over an old, dust-free population is also found 
in the 5.8 $\mu$m and particularly in the 8 $\mu$m IRAC bands. The excess at 8 $\mu$m is likely 
due to polycyclic aromatic hydrocarbons (PAH) transiently heated by UV photons from hot stars.  
Such emission features are commonly found in the spectra of star-forming galaxies and usually accompany a far-infrared spectral peak at $\sim 70 \mu$m.  Adding a nuclear starburst SED model
from \citet{2007A&A...461..445S} of with a total starburst luminosity of $0.95 \times 10^{44}$ erg s$^{-1}$ to the emission from old stars yields an adequate fit (solid line) to the 
combined 2MASS and Spitzer photometry for this galaxy.  Note that PAH features are not expected 
from dust that has been exposed to a harsh X-ray radiation environment, since tiny grains are easily 
disrupted by X-rays (Voit 1992). This excess, if confirmed to be PAH features, is an argument against
the dusty gas originating from condensations from the hot ICM.

The 70 $\mu$m/160 $\mu$m flux ratio is larger than that observed for higher redshift ($z=0.25-0.3$) 
BCGs by Egami et al. (2006).  The observed ratios seen for Abell 1835 and Zwicky 3146 by
Egami et al. (2006) are 0.4-0.6, corresponding to a rest-frame black body 
temperature of $\sim35-40$K. The observed ratio for Abell 2597 is $\sim1.5$, corresponding
to a rest-frame black body temperature of $65-75$K. 
This finding may mean that the temperature of the hot dust 
in Abell 2597 is greater than in higher redshift BCGs that are forming stars more quickly 
than the BCG in Abell 2597.  It may seem puzzling that a galaxy
with a lower star formation rate (SFR) seems to have hotter dust than 
large-SFR galaxies. 
One possible explanation is that galaxies with very low SFRs may have only cool dust (if any), 
those with an intermediate SFR may have warmer dust, and perhaps those with a very high 
SFR may destroy the dust in the star formation regions. The dust in these systems might then be
farther from the heat sources, therefore generating a higher IR luminosity but at a 
lower dust temperature.  Another explanation is suggested by 
the radiative transfer models presented in \citet{2007A&A...461..445S} that 
indicate that the very luminous ($L>10^{12.5}L_\odot$) 
sources are cooler because for a given $A_V$, the dust mass $M_d$ increases
like $R^2$, and the dust temperature scales like $L/M_d$ \citep{2007A&A...461..445S}. 
Also, a large fraction of buried OB stars increases the near-IR flux. Mid-IR spectra, with more than 2-3 
points per spectrum, of these sources
will provide an interesting discriminant between these model SEDs.

Assuming for the moment that the far-IR peak is indeed from star-forming regions, we can compare the implied star-formation rate with that inferred from the H$\alpha$ emission. The diameter of the
H$\alpha$ nebula is about $16\arcsec$ \citep{HBvM1989}, although the high-surface brightness
structure visible in the HST image from \citet{Donahue2000}  
has a similar diameter to that of the radio source ($\sim5-6\arcsec$).
Corrected to $H_0=70$ km s$^{-1}$ Mpc$^{-1}$, \citet{HBvM1989} measure a total H$\alpha$ + [N~II] luminosity
of $3.1 \times 10^{42}$ erg s$^{-1}$, which is $\sim 1.3-1.5 \times 10^{42}$ erg s$^{-1}$ in H$\alpha$ alone. (The
ratio of H$\alpha/$[N~II]6584\AA~ varies in this source between 0.8-1.0, and the [N~II]6548\AA~ line flux is
1/3 that of the 6584\AA~ line.)  A reddening analysis of 5 hydrogen Balmer lines in
the spectrum of Abell 2597 by \citet{1997ApJ...486..242V} optical depth at
H$\beta$ is about 1.2. The correction to total H$\alpha$ is approximately 25\%, 
corresponding to $\sim 1.6-1.8 \times 10^{42}$ erg s$^{-1}$ 
Such an analysis is insensitive to absorption by so-called ``grey'' dust, i.e. a population
of large grains whose absorption is independent of wavelength. 
This luminosity corresponds to a total star formation rate of $\sim 12-14$ M$_\odot$ yr$^{-1}$ using
the conversion from \citep{Kennicutt1998}. 
This rate is somewhat larger than the $4 M_\odot \, {\rm yr}^{-1}$ indicated by the far-IR emission but could be an overestimate if any of the H$\alpha$ arises from processes other than star formation (e.g. AGN, shocks).

The OVI detection reported by \citet{2001ApJ...560..187O} suggests a gas cooling rate $20\pm5$ solar masses per year in the central 26 kpc (quantities converted to $H_0=70$ km s$^{-1}$ Mpc$^{-1}$). This quantity is higher than the  star formation rate inferred from Spitzer data by a factor of $\sim 5\pm3$ but  
is consistent with the local cooling rate of $\sim 20~ \rm{M}_\odot ~\rm{yr}^{-1}$ inferred from X-ray spectra by  \citet{2005MNRAS.358..585M}.  

An alternative source of energy for both H$\alpha$ and far-IR emission emission 
is conduction \citep{1989ApJ...345..153S}.  If conduction is not completely suppressed by 
magnetic fields, it must transfer at least some energy from the X-ray gas to the nebula 
and the cooler dusty gas associated with it.  If we assume saturated heat conduction \citep{1977ApJ...211..135C} from the surrounding $kT=4$ keV ICM 
through a spherical surface of radius $r=10 r_{10}$ kpc, we obtain a maximum heating rate
of a few times $10^{43} r_{10}^2$ erg s$^{-1}$, which represents a significant fraction of the mid-IR
luminosity of $10^{44}$ erg s$^{-1}$.  However, many factors could reduce the conductive
heating rate, such as
suppression of conduction via tangled magnetic fields or the deposition of energy into ionized
nebular gas instead of into the grains.  Additional theoretical work is needed to explore more quantitatively the effect of the surrounding hot ICM on the dusty clouds responsible for the far-IR
emission.


In summary, the UV observations of \cite{ODea2004} place a lower limit on the
star-formation rate because extinction corrections revise the UV rate upwards; the H$\alpha$ luminosity
provides only an approximate limit since other processes can generate H$\alpha$, and
H$\alpha$ can be attenuated by grey absorption. 
These far-IR infrared observations place the best upper limit on 
the obscured star formation, because alternative contributions (such
as conduction from the ICM) would also revise the inferred star formation rate downward. 
Because we have estimated above that 
conduction could in principle provide up to $\sim 50$\% of the  IR luminosity, 
 the uncertainty of the IR-inferred star formation rate is at least a factor of two, 
and we require better theoretical models and more detailed spectra to improve our estimates.

\section{Conclusions}

We have detected mid-infrared emission from the central galaxy of Abell 2597 with the Spitzer 
Space Telescope, from 3.6 - 160 $\mu$m. This galaxy has the luminosity and spectral shape of a LIRG embedded
in a giant elliptical galaxy. The far infrared luminosity rivals that of the local X-ray luminosity.
We have constructed a broad-band spectrum of the 
galaxy, which is most simply interpreted as a dust-enshrouded stellar population forming stars at
a rate $\sim 4 M_\odot \, {\rm yr}^{-1}$ solar masses per year inside the central $35\arcsec$ ($\sim 54$ kpc).
This estimate is consistent within a factor of two of 
estimates from optical, H$\alpha$, and ultraviolet observations. We cannot, however, rule out 
additional heating of the dust by electron thermal conduction from the hot gas, and we
suggest the development of more quantitative physical models of this process. The presence of UV continuum light argues
in favor of a substantial fraction of the mid-IR flux being associated with star formation.
The star formation rate inferred from the mid-IR emission is somewhat lower than the cooling rate
inferred from OVI and recent X-ray observations, but we cannot rule out the hypothesis that 
cooling gas may feed the star formation implied by the UV, optical, and mid-IR data.
 
Finally, we have have been able to model the broadband infrared SED of the 
BCG with a basic giant elliptical template and
a standard nuclear starburst model from  \citet{2007A&A...461..445S}, including PAH features. 
We therefore have no evidence, based on these data, that the dusty gas condensed from the hot ICM. Further spectroscopic detail is required to test the hypothesis that PAHs are responsible for the mid-IR excess. 

\acknowledgements

Support for Donahue was provided by a NASA Spitzer contract (JPL 1268128) 
and a NASA LTSA grant (NASA NNG-05GD82G).
MD acknowledges useful discussions about Spitzer photometry 
with the Spitzer helpdesk personnel and with Dr. Grant Tremblay.
WBS acknowledges support from NASA Spitzer contract JPL 1269604.


\begin{thebibliography}{41}
\expandafter\ifx\csname natexlab\endcsname\relax\def\natexlab#1{#1}\fi

\bibitem[{{Barnes} \& {Nulsen}(2003)}]{2003MNRAS.343..315B}
{Barnes}, D.~G., \& {Nulsen}, P.~E.~J. 2003, \mnras, 343, 315

\bibitem[{{Best} {et~al.}(2006){Best}, {Kaiser}, {Heckman}, \&
  {Kauffmann}}]{2006MNRAS.368L..67B}
{Best}, P.~N., {Kaiser}, C.~R., {Heckman}, T.~M., \& {Kauffmann}, G. 2006,
  \mnras, 368, L67

\bibitem[{{Binney}(2004)}]{2004MNRAS.347.1093B}
{Binney}, J. 2004, \mnras, 347, 1093

\bibitem[{{Bolzonella} {et~al.}(2000){Bolzonella}, {Miralles}, \&
  {Pell{\'o}}}]{2000A&A...363..476B}
{Bolzonella}, M., {Miralles}, J.-M., \& {Pell{\'o}}, R. 2000, \aap, 363, 476

\bibitem[{{Bregman} {et~al.}(2006){Bregman}, {Fabian}, {Miller}, \&
  {Irwin}}]{2006ApJ...642..746B}
{Bregman}, J.~N., {Fabian}, A.~C., {Miller}, E.~D., \& {Irwin}, J.~A. 2006,
  \apj, 642, 746

\bibitem[{{Cardiel} {et~al.}(1998){Cardiel}, {Gorgas}, \&
  {Aragon-Salamanca}}]{Cardiel1998}
{Cardiel}, N., {Gorgas}, J., \& {Aragon-Salamanca}, A. 1998, \mnras, 298, 977

\bibitem[{{Cohen} {et~al.}(2003){Cohen}, {Wheaton}, \&
  {Megeath}}]{2003AJ....126.1090C}
{Cohen}, M., {Wheaton}, W.~A., \& {Megeath}, S.~T. 2003, \aj, 126, 1090

\bibitem[{{Cowie} \& {McKee}(1977)}]{1977ApJ...211..135C}
{Cowie}, L.~L., \& {McKee}, C.~F. 1977, \apj, 211, 135

\bibitem[{{Donahue} {et~al.}(2006){Donahue}, {Horner}, {Cavagnolo}, \&
  {Voit}}]{2006ApJ...643..730D}
{Donahue}, M., {Horner}, D.~J., {Cavagnolo}, K.~W., \& {Voit}, G.~M. 2006,
  \apj, 643, 730

\bibitem[{{Donahue} {et~al.}(2000){Donahue}, {Mack}, {Voit}, {Sparks},
  {Elston}, \& {Maloney}}]{Donahue2000}
{Donahue}, M., {Mack}, J., {Voit}, G.~M., {Sparks}, W., {Elston}, R., \&
  {Maloney}, P.~R. 2000, \apj, 545, 670

\bibitem[{{Donahue} {et~al.}(2005){Donahue}, {Voit}, {O'Dea}, {Baum}, \&
  {Sparks}}]{Donahue2005A}
{Donahue}, M., {Voit}, G.~M., {O'Dea}, C.~P., {Baum}, S.~A., \& {Sparks}, W.~B.
  2005, \apjl, 630, L13

\bibitem[{{Edge}(2001{\natexlab{a}})}]{2001MNRAS.328..762E}
{Edge}, A.~C. 2001{\natexlab{a}}, \mnras, 328, 762

\bibitem[{{Edge}(2001{\natexlab{b}})}]{Edge2001}
---. 2001{\natexlab{b}}, \mnras, 328, 762

\bibitem[{{Edge} \& {Frayer}(2003)}]{2003ApJ...594L..13E}
{Edge}, A.~C., \& {Frayer}, D.~T. 2003, \apjl, 594, L13

\bibitem[{{Egami} {et~al.}(2006{\natexlab{a}}){Egami}, {Misselt}, {Rieke},
  {Wise}, {Neugebauer}, {Kneib}, {Le Floc'h}, {Smith}, {Blaylock}, {Dole},
  {Frayer}, {Huang}, {Krause}, {Papovich}, {P{\'e}rez-Gonz{\'a}lez}, \&
  {Rigby}}]{2006ApJ...647..922E}
{Egami}, E., {Misselt}, K.~A., {Rieke}, G.~H., {Wise}, M.~W., {Neugebauer}, G.,
  {Kneib}, J.-P., {Le Floc'h}, E., {Smith}, G.~P., {Blaylock}, M., {Dole}, H.,
  {Frayer}, D.~T., {Huang}, J.-S., {Krause}, O., {Papovich}, C.,
  {P{\'e}rez-Gonz{\'a}lez}, P.~G., \& {Rigby}, J.~R. 2006{\natexlab{a}}, \apj,
  647, 922

\bibitem[{{Egami} {et~al.}(2006{\natexlab{b}}){Egami}, {Rieke}, {Fadda}, \&
  {Hines}}]{2006ApJ...652L..21E}
{Egami}, E., {Rieke}, G.~H., {Fadda}, D., \& {Hines}, D.~C. 2006{\natexlab{b}},
  \apjl, 652, L21

\bibitem[{{Elston} \& {Maloney}(1994)}]{1994iaan.conf..169E}
{Elston}, R., \& {Maloney}, P. 1994, in ASSL Vol. 190: Astronomy with Arrays,
  The Next Generation, ed. I.~S. {McLean}, 169--+

\bibitem[{{Fabian}(1994)}]{1994ARA&A..32..277F}
{Fabian}, A.~C. 1994, \araa, 32, 277

\bibitem[{{Fazio} {et~al.}(2004){Fazio}, {Hora}, {Allen}, {Ashby}, {Barmby},
  {Deutsch}, {Huang}, {Kleiner}, {Marengo}, {Megeath}, {Melnick}, {Pahre},
  {Patten}, {Polizotti}, {Smith}, {Taylor}, {Wang}, {Willner}, {Hoffmann},
  {Pipher}, {Forrest}, {McMurty}, {McCreight}, {McKelvey}, {McMurray}, {Koch},
  {Moseley}, {Arendt}, {Mentzell}, {Marx}, {Losch}, {Mayman}, {Eichhorn},
  {Krebs}, {Jhabvala}, {Gezari}, {Fixsen}, {Flores}, {Shakoorzadeh}, {Jungo},
  {Hakun}, {Workman}, {Karpati}, {Kichak}, {Whitley}, {Mann}, {Tollestrup},
  {Eisenhardt}, {Stern}, {Gorjian}, {Bhattacharya}, {Carey}, {Nelson},
  {Glaccum}, {Lacy}, {Lowrance}, {Laine}, {Reach}, {Stauffer}, {Surace},
  {Wilson}, {Wright}, {Hoffman}, {Domingo}, \& {Cohen}}]{2004ApJS..154...10F}
{Fazio}, G.~G., {et~al.}
  2004, \apjs, 154, 10

\bibitem[{{Gordon} {et~al.}(2005){Gordon}, {Rieke}, {Engelbracht}, {Muzerolle},
  {Stansberry}, {Misselt}, {Morrison}, {Cadien}, {Young}, {Dole}, {Kelly},
  {Alonso-Herrero}, {Egami}, {Su}, {Papovich}, {Smith}, {Hines}, {Rieke},
  {Blaylock}, {P{\'e}rez-Gonz{\'a}lez}, {Le Floc'h}, {Hinz}, {Latter},
  {Hesselroth}, {Frayer}, {Noriega-Crespo}, {Masci}, {Padgett}, {Smylie}, \&
  {Haegel}}]{2005PASP..117..503G}
{Gordon}, K.~D., {et~al.} 2005, \pasp, 117, 503

\bibitem[{{Heckman} {et~al.}(1989{\natexlab{a}}){Heckman}, {Baum}, {van
  Breugel}, \& {McCarthy}}]{1989ApJ...338...48H}
{Heckman}, T.~M., {Baum}, S.~A., {van Breugel}, W.~J.~M., \& {McCarthy}, P.
  1989{\natexlab{a}}, \apj, 338, 48

\bibitem[{{Heckman} {et~al.}(1989{\natexlab{b}}){Heckman}, {Baum}, {van
  Breugel}, \& {McCarthy}}]{HBvM1989}
---. 1989{\natexlab{b}}, \apj, 338, 48

\bibitem[{{Jaffe} {et~al.}(2005){Jaffe}, {Bremer}, \&
  {Baker}}]{2005MNRAS.360..748J}
{Jaffe}, W., {Bremer}, M.~N., \& {Baker}, K. 2005, \mnras, 360, 748

\bibitem[{{Jaffe} {et~al.}(2001){Jaffe}, {Bremer}, \& {van der
  Werf}}]{2001MNRAS.324..443J}
{Jaffe}, W., {Bremer}, M.~N., \& {van der Werf}, P.~P. 2001, \mnras, 324, 443

\bibitem[{{Jarrett} {et~al.}(2000){Jarrett}, {Chester}, {Cutri}, {Schneider},
  {Skrutskie}, \& {Huchra}}]{2000AJ....119.2498J}
{Jarrett}, T.~H., {Chester}, T., {Cutri}, R., {Schneider}, S., {Skrutskie}, M.,
  \& {Huchra}, J.~P. 2000, \aj, 119, 2498

\bibitem[{{Kennicutt}(1998)}]{Kennicutt1998}
{Kennicutt}, Jr., R.~C. 1998, \apj, 498, 541

\bibitem[{{Koekemoer} {et~al.}(1999){Koekemoer}, {O'Dea}, {Sarazin},
  {McNamara}, {Donahue}, {Voit}, {Baum}, \& {Gallimore}}]{Koekemoer1999}
{Koekemoer}, A.~M., {O'Dea}, C.~P., {Sarazin}, C.~L., {McNamara}, B.~R.,
  {Donahue}, M., {Voit}, G.~M., {Baum}, S.~A., \& {Gallimore}, J.~F. 1999,
  \apj, 525, 621

\bibitem[{{McNamara} {et~al.}(1999){McNamara}, {Jannuzi}, {Sarazin}, {Elston},
  \& {Wise}}]{1999ApJ...518..167M}
{McNamara}, B.~R., {Jannuzi}, B.~T., {Sarazin}, C.~L., {Elston}, R., \& {Wise},
  M. 1999, \apj, 518, 167

\bibitem[{{McNamara} {et~al.}(2001{\natexlab{a}}){McNamara}, {Wise}, {Nulsen},
  {David}, {Carilli}, {Sarazin}, {O'Dea}, {Houck}, {Donahue}, {Baum}, {Voit},
  {O'Connell}, \& {Koekemoer}}]{2001ApJ...562L.149M}
{McNamara}, B.~R., {Wise}, M.~W., {Nulsen}, P.~E.~J., {David}, L.~P.,
  {Carilli}, C.~L., {Sarazin}, C.~L., {O'Dea}, C.~P., {Houck}, J., {Donahue},
  M., {Baum}, S., {Voit}, M., {O'Connell}, R.~W., \& {Koekemoer}, A.
  2001{\natexlab{a}}, \apjl, 562, L149

\bibitem[{{McNamara} {et~al.}(2001{\natexlab{b}}){McNamara}, {Wise}, {Nulsen},
  {David}, {Carilli}, {Sarazin}, {O'Dea}, {Houck}, {Donahue}, {Baum}, {Voit},
  {O'Connell}, \& {Koekemoer}}]{McNamaraA2597_2001}
---. 2001{\natexlab{b}}, \apjl, 562, L149

\bibitem[{{Morris} \& {Fabian}(2005)}]{2005MNRAS.358..585M}
{Morris}, R.~G., \& {Fabian}, A.~C. 2005, \mnras, 358, 585

\bibitem[{{O'Dea} {et~al.}(2004){O'Dea}, {Baum}, {Mack}, {Koekemoer}, \&
  {Laor}}]{ODea2004}
{O'Dea}, C.~P., {Baum}, S.~A., {Mack}, J., {Koekemoer}, A.~M., \& {Laor}, A.
  2004, \apj, 612, 131

\bibitem[{{Oegerle} {et~al.}(2001){Oegerle}, {Cowie}, {Davidsen}, {Hu},
  {Hutchings}, {Murphy}, {Sembach}, \& {Woodgate}}]{2001ApJ...560..187O}
{Oegerle}, W.~R., {Cowie}, L., {Davidsen}, A., {Hu}, E., {Hutchings}, J.,
  {Murphy}, E., {Sembach}, K., \& {Woodgate}, B. 2001, \apj, 560, 187

\bibitem[{{Peterson} {et~al.}(2001){Peterson}, {Paerels}, {Kaastra}, {Arnaud},
  {Reiprich}, {Fabian}, {Mushotzky}, {Jernigan}, \&
  {Sakelliou}}]{2001A&A...365L.104P}
{Peterson}, J.~R., {Paerels}, F.~B.~S., {Kaastra}, J.~S., {Arnaud}, M.,
  {Reiprich}, T.~H., {Fabian}, A.~C., {Mushotzky}, R.~F., {Jernigan}, J.~G., \&
  {Sakelliou}, I. 2001, \aap, 365, L104

\bibitem[{{Reach} {et~al.}(2005){Reach}, {Megeath}, {Cohen}, {Hora}, {Carey},
  {Surace}, {Willner}, {Barmby}, {Wilson}, {Glaccum}, {Lowrance}, {Marengo}, \&
  {Fazio}}]{2005PASP..117..978R}
{Reach}, W.~T., {Megeath}, S.~T., {Cohen}, M., {Hora}, J., {Carey}, S.,
  {Surace}, J., {Willner}, S.~P., {Barmby}, P., {Wilson}, G., {Glaccum}, W.,
  {Lowrance}, P., {Marengo}, M., \& {Fazio}, G.~G. 2005, \pasp, 117, 978

\bibitem[{{Salom{\'e}} {et~al.}(2006){Salom{\'e}}, {Combes}, {Edge},
  {Crawford}, {Erlund}, {Fabian}, {Hatch}, {Johnstone}, {Sanders}, \&
  {Wilman}}]{2006A&A...454..437S}
{Salom{\'e}}, P., {Combes}, F., {Edge}, A.~C., {Crawford}, C., {Erlund}, M.,
  {Fabian}, A.~C., {Hatch}, N.~A., {Johnstone}, R.~M., {Sanders}, J.~S., \&
  {Wilman}, R.~J. 2006, \aap, 454, 437

\bibitem[{{Schlegel} {et~al.}(1998){Schlegel}, {Finkbeiner}, \&
  {Davis}}]{1998ApJ...500..525S}
{Schlegel}, D.~J., {Finkbeiner}, D.~P., \& {Davis}, M. 1998, \apj, 500, 525

\bibitem[{{Siebenmorgen} \& {Kr{\"u}gel}(2007)}]{2007A&A...461..445S}
{Siebenmorgen}, R., \& {Kr{\"u}gel}, E. 2007, \aap, 461, 445

\bibitem[{{Sparks} {et~al.}(1989){Sparks}, {Macchetto}, \&
  {Golombek}}]{1989ApJ...345..153S}
{Sparks}, W.~B., {Macchetto}, F., \& {Golombek}, D. 1989, \apj, 345, 153

\bibitem[{{Voit} \& {Donahue}(1997{\natexlab{a}})}]{1997ApJ...486..242V}
{Voit}, G.~M., \& {Donahue}, M. 1997{\natexlab{a}}, \apj, 486, 242

\bibitem[{{Voit} \& {Donahue}(1997{\natexlab{b}})}]{VD1997}
---. 1997{\natexlab{b}}, \apj, 486, 242

\end{thebibliography}

\clearpage

\begin{table}
\caption{Infrared Fluxes for the BCG in Abell 2597 \label{table:fluxes}}
\begin{center}
\begin{tabular}{lcccccccccc} \tableline
Band ($\mu$m)              & 1.235 &  1.662  & 2.159 & 3.6 & 4.5 & 5.8 & 8.0          & 24   & 70   & 160 \\ \tableline
Flux (Unc. mJy) \tablenotemark{1}   & 9.7 &  9.3  & 9.64 & 6.0 & 4.0 & 3.1 & 2.7      & 2.1  & 89   &  35 (52) \\
Error  (mJy)    \tablenotemark{2}   & 0.6 &  1.1  & 0.96 & 0.8 & 0.2 & 0.2 & 0.01     & 0.2  &  4  &    2 (3) \\ \tableline
PRF Flux (mJy) \tablenotemark{3}    &     &       &      &     &     &     &          & 2.09  & 86 &   57.0 \\
Error (mJy) \tablenotemark{2}       &     &       &      &     &     &     &          & 0.06 &  1  &    1.6 \\ \tableline
\end{tabular}
\end{center}
\tablenotetext{1}{Aperture photometry. 
Cataloged 2MASS total magnitudes were measured with a large aperture $r=25.35\arcsec$, IRAC and MIPS with $25\arcsec$. MIPS 
160 micron aperture fluxes with $r=35\arcsec$ aperture are also reported. Based on test convolutions with 
the 160 micron PRF, the aperture correction for a $35\arcsec$ radius aperture is 1.5, corrected flux shown in parentheses. }
\tablenotetext{2}{The Spitzer measurement 
uncertainties quoted here are dominated by background subtraction. IRAC absolute photometry is good to about 3\% for
point sources \citep{2005PASP..117..978R}.
From the MIPS data handbook, MIPS photometric calibration 
uncertainties of 10\% for 24 $\mu$m and 20\% for 70 and 160 $\mu$m apply.}
\tablenotetext{3}{Point response fit (PRF) photometry for MIPS. No aperture correction is needed for the
2MASS or IRAC photometry for this size of aperture and a galaxy of this angular size. 
To first approximation, the MIPS sources are
point-like, so a PRF fit provides a reasonable estimate of the total flux.}
\end{table}

\clearpage

\end{document}